\documentclass[usenatbib]{mn2e}
\usepackage{graphicx}
\usepackage[usenames]{color}
\usepackage{euscript}
\usepackage{amssymb}
\usepackage{journals}

\definecolor{grey}{rgb}{0.5,0.6,0.7}

\title[Stellar Populations and Dark Matter in UCDs]{Stellar Population
Constraints on the Dark Matter Content and Origin of Ultra-Compact Dwarf 
Galaxies\thanks{Based on the archival ESO VLT data (programmes 074.A-0756,
074.A-0508) available through http://archive.eso.org/}}
\author[I. Chilingarian, V. Cayatte \&
  G. Bergond]{Igor V. Chilingarian$^{1,2}$\thanks{E-mail:
    Igor.Chilingarian@obspm.fr}, V\'eronique Cayatte$^{3}$ 
    and Gilles Bergond$^{4}$\\
$^{1}$Observatoire de Paris-Meudon, LERMA, UMR~8112, 61 Av. de
  l'Observatoire, 75014 Paris, France\\
$^{2}$Sternberg Astronomical Institute, Moscow State University, 13 Universitetski prospect, 119992, Moscow, Russia\\
$^{3}$Observatoire de Paris-Meudon, LUTH, UMR~8102, 5 pl. Jules Janssen, 92195
Meudon, France\\
$^{4}$Instituto de Astrof\'isica de Andaluc\'ia (IAA/CSIC), C/ Camino Bajo de Hu\'etor 50, 18008 Granada, Espa\~na
}
\begin{document}

\date{Accepted 2008 August 4. Received 2008 June 23; in original form 2008 May 16}

\pagerange{\pageref{firstpage}--\pageref{lastpage}} \pubyear{2008}

\maketitle

\label{firstpage}

\begin{abstract} 
We analyse intermediate-resolution VLT FLAMES/Giraffe spectra of six
ultra-compact dwarf (UCD) galaxies in the Fornax cluster. We obtained
velocity dispersions and stellar population properties by full spectral
fitting against {\sc pegase.hr} models. Objects span a large range of
metallicities (--0.95 to --0.23~dex), 4 of them are older than 8~Gyr.
Comparison of the stellar and dynamical masses suggests that UCDs have
little dark matter at best. For one object, UCD3, the Salpeter initial
mass function (IMF) results in the stellar mass significantly exceeding the
dynamical one, whereas for the Kroupa IMF the values coincide.
Although, this object may have peculiar dynamics or/and stellar populations,
the Kroupa IMF seems more realistic. We find that UCDs lie well above the
metallicity--luminosity relation of early-type galaxies. The same behaviour
is demonstrated by some of the massive Milky Way globular clusters,
known to contain composite stellar populations. Our results support two following
UCD formation scenarii: (1) tidal stripping of nucleated dwarf elliptical
galaxies; (2) formation of tidal superclusters in galaxy mergers. We
also discuss some of the alternative channels of the UCD formation binding
them to globular clusters.
\end{abstract}

\begin{keywords}
galaxies: dwarf -- galaxies: elliptical and lenticular, cD -- 
galaxies: evolution -- galaxies: stellar content --
galaxies: kinematics and dynamics % -- galaxies: star clusters
\end{keywords}

\section{Introduction} 

A new class of compact dwarf galaxies, called UCDs, was discovered a decade
ago in the Fornax cluster \citep{Hilker+99,Drinkwater+00,PDGJ01}.  These
objects, being brighter and much larger than globular clusters (GCs), but
still far below luminosities and sizes of both dwarf elliptical (dE) and
compact elliptical (cE) galaxies, fill an empty region on the Fundamental
Plane \citep{DD87}.  Their origin still remains a matter of debate; several
alternatives are considered: (1) UCDs are the result of the evolution of
primordial density fluctuations \citep{PDGJ01}; (2) they have been formed
through mergers of GCs or simply represent the extreme high-luminosity end
of the GC luminosity function \citep{MHI02}; (3) UCDs are nuclei of tidally
stripped (``threshed'') nucleated dE (dE,N) galaxies \citep{BCDS03} or dE,Ns
with very low surface brightness; (4) UCDs are created as tidal
superclusters during major mergers of galaxies \citep{FK05,KJB06}.

Mass-to-light ratios of UCDs vary quite significantly
\citep{Drinkwater+03,Hasegan+05,Hilker+07}, suggesting the presence of dark
matter in some of them. \cite{Hasegan+05} propose to use $M/L$ ratio (i.e.
presence of dark matter) as a criterion to distinguish between ``UCD
galaxies'' and massive GCs.  Developing this idea, we conclude that the
presence of dark matter in a compact stellar system rejects two
formation scenarii -- in this case UCDs cannot be GCs neither they can be
created as tidal superclusters during galaxy mergers. On the other hand, if
the stellar population is not old and metal-poor, the primordial density
fluctuation scenario will become implausible, leaving the only channel of
UCD formation to be the tidal stripping of dE,Ns.

Stellar population analysis may also help to choose the formation scenario.
Presently published data \citep{MHIJ06,EGDH07} based on the analysis of
absorption line strengths (Lick indices, \citealp{WFGB94}) suggest that UCDs
are old and rather metal-poor.

In this paper, we present stellar population parameters for 6 Fornax cluster
UCDs, compare them with dE,N nuclei, and derive stellar masses to check for
the presence of dark matter assuming different stellar IMFs.

\section{Data: Sources, Reduction, Analysis}
We have used the data 
obtained in the courses of two 
independent studies
of compact stellar systems in the Fornax cluster by G. Bergond et al.
(program 074.A-0756) and M. Drinkwater et al. (program
074.A-0508). Both datasets are publicly available through the ESO Data
Archive.
The data have been obtained with
the ESO Very Large Telescope using the FLAMES/Giraffe spectrograph
\citep{Pasquini+02} in the
multi-object ``MEDUSA'' mode (130 fibres in a 25~arcmin
circular field of view), in the LR04 setup giving a resolving power 
$R\approx6300$ in the wavelength range 5010--5831~\AA\ (dispersion 0.2
\AA\,pix$^{-1}$, $\sigma_{\rm{inst}} \approx 18$ km\,s$^{-1}$), and reduced in exactly the same way as the data 
for the Abell~496 cluster as described in \cite{Chil08A496}. The 1.2~arcsec-wide FLAMES/Giraffe 
fibres corresponding to a spatial size of about 110~pc at the Fornax distance 
(19 Mpc), are significantly larger than the typical effective radii of UCDs \citep{EGDH07}. 
Therefore, our stellar population and velocity dispersion measurements
referenced below should be considered as global values (however, see
discussion about aperture corrections in \citealp{Mieske+08}).

Observations of the central part of the Fornax cluster produced about 900
individual spectra (see details on observations in \citealp{Bergond+07} and
\citealp{Firth+07}). We inspected them visually to identify those having
reasonable signal-to-noise ratios and also to take out background galaxies;
spectra of the objects common to the two studies have been co-added.

We ended up with a list including about 40 spectra of foreground Milky Way
stars and members of the Fornax cluster of different nature: GCs, UCDs, dE
and non-dwarf galaxies. In this paper we analyse six UCDs and two dE,N
nuclei (FCC~182 and FCC~266), while the brightest GCs and other Fornax
cluster members will be presented in detail in the forthcoming paper. Our
sample is presented in Table~\ref{tabucdlist}.

\begin{table}
\caption{Final sample of 
UCDs and dEs. (2) and (3) give
identification according to Bergond et al. (2007) and Evstigneeva et al.
(2007), (4)--(7) provide the number of 
individual exposures and the total exposure times (seconds) in the two 
observational programmes (B and D indices are for Bergond et al. and
Drinkwater et al.), (8) lists approximate signal-to-noise ratio
per pixel at $\lambda = 5300$~\AA for the combined spectra.
\label{tabucdlist}
}

\begin{tabular}{llcccccc}
\hline
$n$ & ID$_{\rm B}$ & ID$_{\rm E}$ & $n_{\rm B}$ & $t_{\rm B}$, s &  $n_{\rm D}$
& $t_{\rm D}$, s & S/N\\
\hline
1 & ucd257.5 & {\bf UCD1} & 3 & 10200 & 3 & 7200 & 19 \\
2 & ucdA & {\bf UCD2} & 3 & 10200 & - & - & 9 \\
3 &  & {\bf UCD3} & - & - & 3 & 6600 & 15 \\ 
4 & ucdB & {\bf UCD4} & 2 & 7200 & 3 & 6600 & 13 \\
5 &  & {\bf UCD5} & - & - & 3 & 6000 & 9 \\
6 & {\bf ucd329.7} &  & 3 & 11200 & - & - & 15 \\
\hline
7 & {\bf FCC182} & & 3 & 10200 & - & - & 40 \\
8 & {\bf FCC266} & & 3 & 9704 & - & - & 17 \\
\hline
\end{tabular}
\end{table}

We have fit the high-resolution {\sc pegase.hr} (Le Borgne et al. 2004)
simple stellar population (SSP) models against the observational data using
the {\sc NBursts} full spectral fitting technique \citep{CPSA07,CPSK07}.
The fitting algorithm works as follows: (1) a grid of SSP spectra with a fixed
set of ages (nearly logarithmically spaced from 20~Myr to 18~Gyr) and
metallicities (from $-$2.0 to $+$0.5~dex) is convolved with the instrumental
response of FLAMES/Giraffe as explained in Section~4.1 of \cite{CPSA07}; (2)
a non-linear least square fitting against an observed spectrum is done for a
template picked up from the pre-convolved SSP grid using 2D-spline
interpolation on $\log t$ and $Z$, broadened according to the
line-of-sight velocity distribution (LOSVD) parametrised by $v$ and $\sigma$ and
multiplied pixel-by-pixel by the $n^{\rm{th}}$ order Legendre polynomial,
resulting in $n + 5$ parameters determined by the non-linear fitting. For
the spectra presented in this paper we used the pure Gaussian
representation of the LOSVD and did not fit the 
$h_3$ and $h_4$ coefficients of the Gauss-Hermite LOSVD parametrization
\citep{vdMF93} often used to perform the dynamical modelling of galaxies 
due to low signal-to-noise ratio and insufficient sampling of
the LOSVD.

The procedure and
input parameters of the fitting (15$^{\rm{th}}$ order multiplicative 
continuum, etc.) were exactly
the same as ones applied to the sample of Abell~496 low-luminosity
early-type galaxies \citep{Chil08A496}, thus we refer to that paper
for all details concerning the spectral fitting. The only difference 
introduced here is that we use SSP models computed for two different
stellar IMFs: \cite{Salpeter55} and \cite{KTG93}. We use the two grids of template 
spectra (Salpeter and Kroupa SSPs hereafter) in a completely independent
way and provide comparison of the results obtained.
 
The key to the precise determination of low velocity dispersions from
absorption-line spectra is the knowledge of the instrumental resolution as a
function of wavelength and fibre number. We found that fibre-to-fibre
variations in the ``MEDUSA'' mode of FLAMES/Giraffe are negligible, while
wavelength dependence has to be taken into account.

\begin{figure}
\includegraphics[width=\hsize]{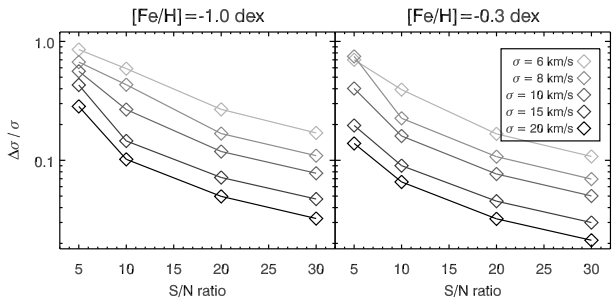}
\caption{Precision of velocity dispersion determinations from
the full spectral fitting of FLAMES/Giraffe-LR04 spectra computed with
Monte-Carlo simulations for different signal-to-noise ratios, internal
velocity dispersion and metallicities. 5 curves correspond to
input velocity dispersions from 6 to 20~km\,s$^{-1}$.\label{figsimsig}}
\end{figure}

We performed Monte-Carlo simulations aimed at studying the precision of
kinematical and stellar population parameters determined by our spectral
fitting technique for objects having very low intrinsic velocity
dispersions, close to or smaller than the instrumental resolution of the
spectrograph. We used two {\sc pegase.hr} SSP models for the age of 10~Gyr and
[Fe/H] $=-1.0$ and $-$0.3~dex, and broadened them using the wavelength-dependent
information about the spectral line spread of FLAMES/Giraffe in the LR04
setup. Then we generated sets of mock data (20 realisations for
every parameter set) for signal-to-noise ratios of 5, 10, 20, and 30 and
internal velocity dispersions of 6, 8, 10, 15, and 20~km\,s$^{-1}$, resulting in 800
mock spectra, which then were fit using the {\sc NBursts} code. The results
are presented in Fig~\ref{figsimsig}.

Our simulations clearly demonstrate that: (1) FLAMES/Giraffe-LR04 is
sufficient to measure internal velocity dispersion down to 8--10~km\,s$^{-1}$ at a
signal-to-noise ratio of 20 with a precision of 10--15 per cent even for
metal-poor ([Fe/H] $=-1.0$) objects; (2) for metal-rich ([Fe/H] $=-0.3$) objects
we reach twice higher precision of internal velocity dispersion measurements
compared to metal-poor ones. We notice that uncertainties of age,
metallicity and radial velocity determinations returned by the least square
fitting are consistent with the results of the Monte-Carlo simulations.

\section{Results}
In Table~\ref{tabresucd} we present the values of the (heliocentric) radial velocities,
velocity dispersions, SSP-equivalent ages and metallicities, and
$B$-band stellar mass-to-light ratios for six UCD galaxies and two dE,N
nuclei computed for the two IMFs mentioned above. We compare our results with the literature for some of the objects.
Absolute magnitudes for 
UCD 1 to 5 are taken from Evstigneeva
et al. (2006) and converted into the $B$ band, for ucd329.7 values are from
Bergond et al. (in prep.), and for FCC~182 and FCC~266 from \cite{KDG03}. 
Velocity dispersion values ($\sigma_{\rm{lit}}$) for UCD 1 to
5 are ``adopted global velocity dispersions'' from Table~6 of
\cite{Hilker+07}; metallicities [Fe/H]$_{\rm{lit}}$ for UCD 2, 3, and 4 
are from \cite{MHIJ06}.

\begin{table*}
\caption{Internal kinematics, stellar populations and 
stellar $B$-band mass-to-light ratios of 6 UCDs and 2 dEs ($n=7,8$)
in the Fornax cluster. Columns (5)--(7) and (8)--(10) are for SSP models
computed with Salpeter and Kroupa et al. (2003) IMF, respectively.
\label{tabresucd}}
\begin{tabular}{lccccccccc|cc}
\hline
$n$ & $M_B$ & $v_{\rm{hel}}$ & $\sigma$ & 
$t_{\rm{Salp.}}$ & [Fe/H]$_{\rm S.}$ & ($M/L$)$_{*B}$ & 
$t_{\rm{Kroupa}}$ & [Fe/H]$_{\rm K.}$ & ($M/L$)$_{*B}$ &
$\sigma_{\rm{lit}}$ & [Fe/H]$_{\rm{lit}}$ \\
  & mag & km\,s$^{-1}$ & km\,s$^{-1}$ & Gyr & dex & Salpeter
& Gyr & dex & Kroupa 
& km\,s$^{-1}$ & dex \\
\hline
1      & -11.39 & 1557$\pm$1  & 29$\pm$1 &  9.1$\pm$2.4  & -0.46$\pm$0.04 &  5.2$\pm$1.3 & 13.1$\pm$3.1 & -0.51$\pm$0.07 &  3.8$\pm$0.9 & 27$\pm$2 & \\
2      & -11.47 & 1230$\pm$1  & 23$\pm$2 &  5.0$\pm$1.7  & -0.24$\pm$0.07 &  3.8$\pm$1.3 &  5.0$\pm$1.9 & -0.23$\pm$0.08 &  2.3$\pm$0.8 & 22$\pm$2 & -0.90 \\
3      & -12.77 & 1500$\pm$1  & 26$\pm$1 & 13.0$\pm$2.5  & -0.23$\pm$0.05 &  8.6$\pm$1.5 & 17.6$\pm$2.7 & -0.22$\pm$0.05 &  6.3$\pm$1.1 & 23$\pm$3 & -0.52 \\
4      & -11.65 & 1889$\pm$1  & 26$\pm$1 &  8.1$\pm$2.4  & -0.67$\pm$0.06 &  4.0$\pm$1.1 &  9.9$\pm$3.7 & -0.68$\pm$0.07 &  2.6$\pm$0.9 & 25$\pm$3 & -0.85 \\
5      & -11.19 & 1280$\pm$2  & 16$\pm$3 &  3.9$\pm$0.9  & -0.95$\pm$0.04 &  1.8$\pm$0.4 &  3.9$\pm$0.9 & -0.93$\pm$0.04 &  1.1$\pm$0.3 & 19$\pm$3 & \\
6      & -10.78 & 1379$\pm$1  & 28$\pm$1 & 11.2$\pm$2.4  & -0.30$\pm$0.04 &  7.1$\pm$1.4 & 14.0$\pm$4.0 & -0.33$\pm$0.08 &  4.8$\pm$1.3 & & \\
\hline
7      & -16.50 & 1700$\pm$1  & 38$\pm$1 &  6.8$\pm$0.5  & -0.10$\pm$0.02 &  5.6$\pm$0.5 & 8.0$\pm$0.7 & -0.13$\pm$0.01 & 3.6$\pm$0.2 &          &                \\
8      & -15.40 & 1551$\pm$1  & 19$\pm$1 &  1.8$\pm$0.3  & -0.24$\pm$0.02 &  1.4$\pm$0.3 & 2.4$\pm$0.4 & -0.30$\pm$0.02 & 1.2$\pm$0.2 & 22$\pm$7$^{1}$ & -0.47$^{2}$\\
\hline
\multicolumn{12}{l}{$^1$\footnotesize{Central $\sigma$ given by S. De Rijcke (priv. comm.) is lower than the global value (44~km\,s$^{-1}$) from \cite{deRijcke+05}.}} \\
\multicolumn{12}{l}{$^2$\footnotesize{Metallicity estimation from the full spectral fitting of the VLT FORS1 spectrum \citep{Michielsen+07}.}}
\end{tabular}
\end{table*}

\begin{figure*}
\includegraphics[angle=90,width=\hsize]{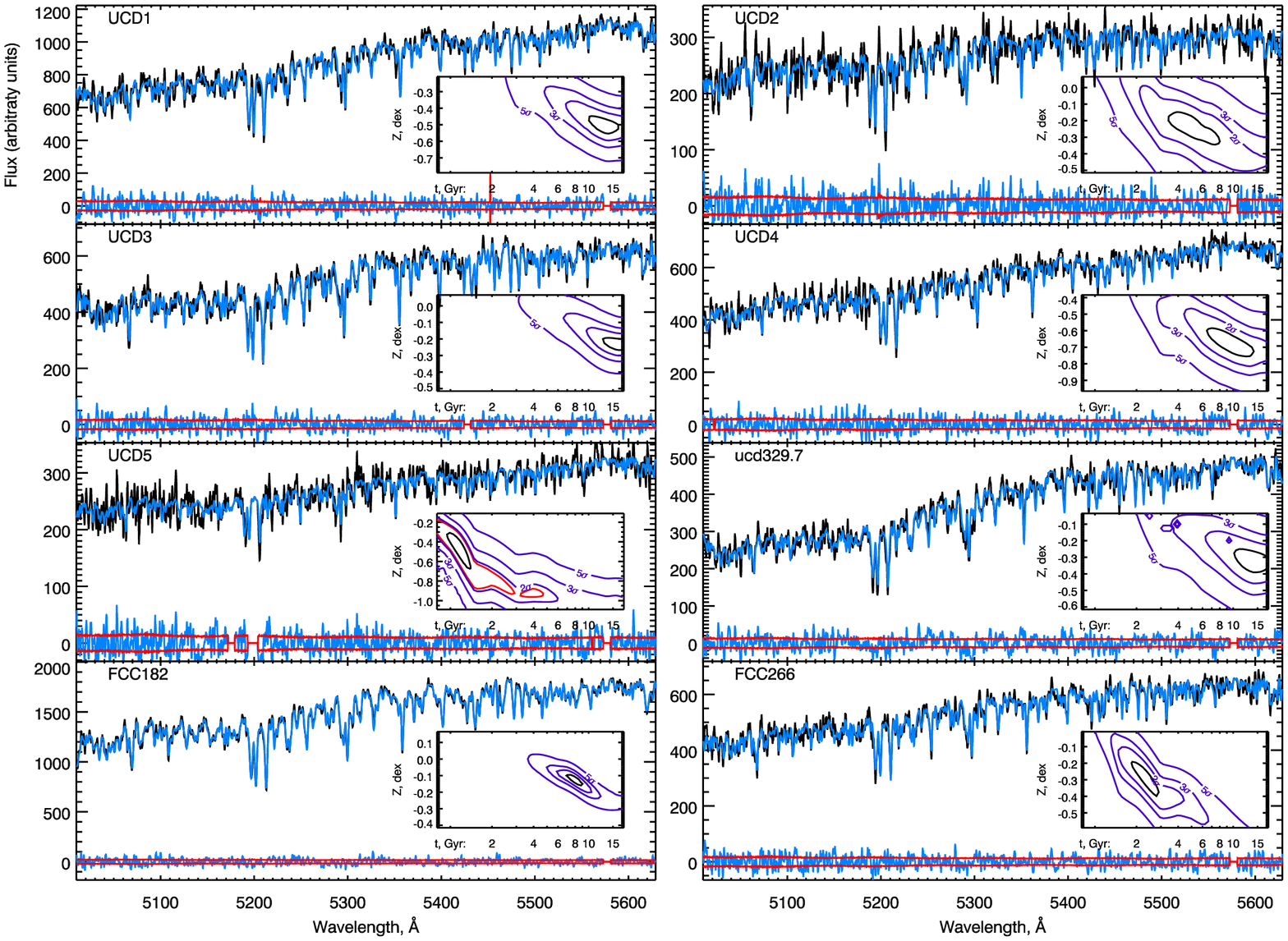}
\caption{FLAMES/Giraffe spectra, their best-fitting templates (Kroupa IMF), 
fitting residuals and confidence levels of the age and metallicity
determinations (inner panels) for 6 UCDs and 2 dE,Ns. 
All graphs are smoothed with the 7~pixel wide box-car for presentation purposes.
\label{figspec}}
\end{figure*}

In Fig.~\ref{figspec} the discussed spectra are displayed together with their
best-fitting {\sc pegase.hr} SSPs. Inner panels show confidence levels of the
age and metallicity determinations for the \cite{KTG93} IMF.

The velocity dispersions for 5 of the  6 UCDs 
are between 23 and
30~km\,s$^{-1}$, which is higher 
than typical values for GCs 
(e.g. \citealp{EGDH07}).

One notices an excellent agreement of our velocity dispersion measurements
for the five UCDs (UCD1--5) and global velocity dispersions from
\cite{Hilker+07} obtained using completely different instrumentation (VLT UVES,
Keck ESI for UCD1) and data analysis technique.

Stellar populations of UCD1, UCD3, UCD4 and ucd329.7 are older than 8~Gyr
and exhibit metallicities between $-$0.67 and $-$0.23~dex. UCD2 has intermediate
age and quite metal-rich stellar population, although with large
uncertainties. Our estimates of metallicities for UCD3, and UCD4 are somewhat
(0.2--0.25) higher than values reported by \cite{MHIJ06}, 
but the discrepancy is even larger ($\approx$0.65~dex) for UCD2.

The UCD5 spectrum does not contain strong absorption lines in the
FLAMES/Giraffe LR04 spectral range, the Mg$b$ triplet is strongly
contaminated by 
a cosmic ray hit. This causes very vague stellar
population parameter determination: 3~$\sigma$ confidence contour remains
open in the age direction, i.e. age is undetermined. The global $\chi^2$
minimum corresponds to a young population ($t=1.2$~Gyr, [Fe/H]
$=-$0.49~dex). There is a 1.7$\sigma$ confidence secondary minimum ($t=3.9$~Gyr,
[Fe/H] $= -$0.95~dex). In order to chose between the two possible solutions for
UCD5 we have reduced and fit its FLAMES/Giraffe spectra obtained in the blue
LR02 setup with the wavelength coverage between 3970 \AA\ and 4545 \AA. Quite
low efficiency of the spectrograph in this setup is well compensated by
the blue colour of the object and the presence of very strong age- and
metallicity-sensitive absorption lines (Ca I ``H'', G-band, H$\gamma$). The
fitting of the LR02 data results in the stellar population parameters
compatible with the secondary minimum of the LR04 setup, therefore we adopt
the secondary LR04 solution, corresponding to the metal-poor intermediate age
population through the rest of the paper. 

Fitting with the Kroupa SSPs results in
slightly older ages for intermediate and old stellar populations, whereas
the metallicities and velocity dispersions remain intact. Mass-to-light
ratios are 40--50 per cent 
lower compared to the Salpeter IMF. This effect
is easy to understand keeping in mind that stellar populations with the
Salpeter IMF contain larger amount of faint red dwarf stars, weakly
contributing to the total light, but increasing the total mass. Therefore,
Kroupa IMF-based SSPs having older ages are required to fit ``red'' spectra.

\section{Discussion}
\subsection{Comparison of Stellar and Dynamical Masses}

Given the stellar population parameters and luminosities, we derive stellar
masses of UCDs in our sample. The stellar mass estimates, computed from the
mass-to-light ratios provided by {\sc pegase.2} \citep{FR97}
for Salpeter and Kroupa et al. IMFs are given in
Table~\ref{tabmlcomp}. In the fourth 
column we provide the corrected
dynamical masses, derived by re-normalising the values of \cite{Hilker+07} 
by our velocity dispersion measurements ($M_{\rm{d, corr}} =
M_{\rm{d}} (\sigma / \sigma_{\rm{lit}})^2$. The fifth and sixth column contain the
dark matter fractions estimated from Salpeter and Kroupa SSPs and corrected dynamical masses.

For UCD1, 2, 4, and 5 the Salpeter SSP stellar masses are
consistent with the dynamical ones within uncertainties, although in average
the stellar masses tend to be lower. The Kroupa et al. IMF decreases them
more resulting in a 40--50 per cent (65 for UCD5) upper limits of the dark
matter content. However, for UCD3 the stellar mass derived using the
Salpeter IMF becomes significantly larger than the dynamical estimate (note
negative dark matter fraction),
whereas Kroupa IMF provides almost a perfect match between the two,
suggesting zero dark matter content. Therefore, if we assume no
object-to-object IMF variation, the observations are more in
favour of the Kroupa et al. IMF.

There is a possibility that the dynamical model of UCD3
used by \cite{Hilker+07} was not correct (for example, (1) the outer
component of the UCD3 is not spherically symmetric or (2) there is
significant rotation not taken into account, or (3) velocity dispersions are
anisotropic), which may lead to an underestimated dynamical mass. At the same
time, we cannot exclude that SSP models do not represent well the spectrum
(e.g. the object contains a metal-rich sub-population, which is not properly
modeled). In this case the stellar mass may be overestimated.

Given the large uncertainties of stellar population parameters and,
consequently, stellar mass-to-light ratios, we cannot give a decisive answer
on a question \emph{``Is there dark matter in UCDs?''} However, the main
conclusion we draw is that \emph{UCDs are not dark matter dominated objects}
and at present level of detection, the dark matter content can be explained
by uncertainties of the measurements and looseness of the models used to
derive dynamical and stellar masses.

\begin{table}
\caption{Comparison of stellar masses of 6 UCDs for 
Salpeter (2) and Kroupa (3) IMFs; and the dynamical masses for 5 
objects (5) from Hilker et al. (2007) corrected using our velocity dispersion
estimations, (6)--(7) the dark matter content (per cent) for the Salpeter
and Kroupa et al. IMF.\label{tabmlcomp}}
\begin{tabular}{cccccc}
\hline
$n$ & $M_{*{\rm{Salp.}}}$ & $M_{*{\rm{Kroupa}}}$ & $M_{\rm{d, corr}}$ &
DM$_{\rm{S.}}$  & DM$_{\rm{K.}}$ \\
 & 10$^7 M_{\odot}$ & 10$^7 M_{\odot}$ & 10$^7 M_{\odot}$ & \% & \% \\
\hline
1 &   2.9$\pm$0.7 &  2.1$\pm$0.5 &  3.7$\pm$0.5  &  20 & 45 \\
2 &   2.3$\pm$0.8 &  1.4$\pm$0.5 &  2.4$\pm$0.3  &   0 & 40 \\
3 &  17.5$\pm$3.0 & 12.8$\pm$2.2 & 12.0$\pm$2.4  &--45 &  0 \\
4 &   2.9$\pm$0.8 &  1.9$\pm$0.6 &  4.0$\pm$1.0  &  30 & 50 \\
5 &   0.9$\pm$0.2 &  0.5$\pm$0.1 &  1.4$\pm$0.5  &  35 & 65 \\
6 &   2.3$\pm$0.4 &  1.6$\pm$0.4 & & & \\
\hline
\end{tabular}
\end{table}

\subsection{Metallicity-$M_B$ and metallicity-$\sigma$ relations}

\begin{figure}
\includegraphics[width=0.95\hsize]{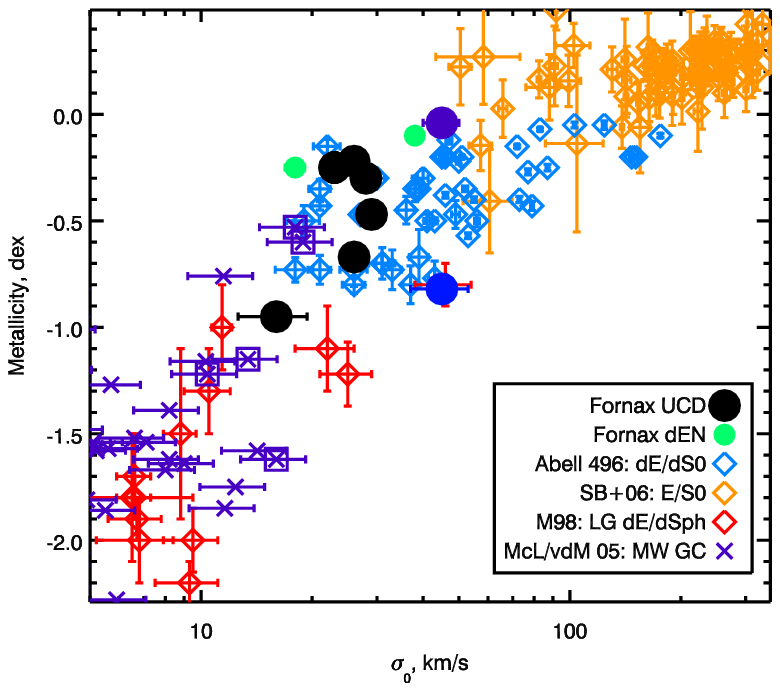}\\
\includegraphics[width=0.95\hsize]{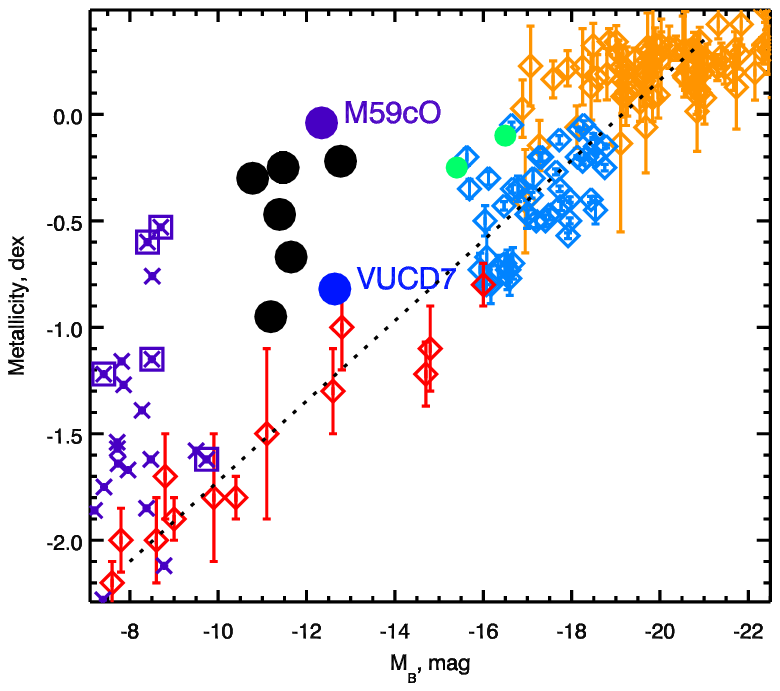}
\caption{Metallicity-velocity dispersion (top) and metallicity-luminosity
(bottom) relations for early-type galaxies, UCDs, and GCs. The data
sources are described in the text. Outlined crosses represent Galactic 
GCs with multi-component stellar populations (see Section 4.2 for details). 
\label{figSLZ}}
\end{figure}

In Fig.~\ref{figSLZ} we plot metallicities versus velocity
dispersions and luminosities of the 6 Fornax cluster UCDs and 2 dE,N nuclei from
our sample. They are compared to: (a) Milky Way GCs from
\cite{MvdM05}, outlined crosses show GCs with direct evidences of multiple
stellar populations revealed by the analysis of colour-magnitude diagrams
\citep{Piotto08}; (b) Local Group dEs and dwarf spheroidals (dSph) from
\cite{Mateo98}; (c) Abell~496 low-luminosity early-type galaxies from
\cite{Chil08A496}; (d) a sample of intermediate-luminosity and bright
early-type galaxies from \cite{SGCG06a,SGCG06c}; two compact
stellar systems in the Virgo cluster with the spectra available in SDSS DR6
\citep{SDSS_DR6}: transitional UCD/cE ``M59cO'' \citep{CM08} and 
VUCD~7, where the data have been processed exactly in
the same way as for M59cO. The aperture size for the Abell~496 dE/dS0 is around
0.8~kpc, so the dE nuclei do not dominate the light, therefore
metallicities and velocity dispersions should be closer to the global than
to the central values, given flat velocity dispersion profiles usually
observed in dEs \citep{SP02, GGvdM03, vZSH04}.

In the metallicity-luminosity relation (Fig.~\ref{figSLZ}, bottom panel)
there is a continuous sequence $Z \varpropto L_B^{0.45}$, spanning over 6
orders of magnitude in luminosity, formed by early-type galaxies: from the
faintest dSph on the left to the brightest cluster ellipticals on the right.
UCD galaxies lie significantly above (0.7--1.0~dex) this sequence, compared
to the brightest Local Volume dSph's, having similar luminosities (see
\citealp{Mateo98} and references therein). In this sense, UCDs are
similar to cE galaxies (e.g. \citealp{Chilingarian+07}) having high
metallicities for their luminosities, which probably represent end-products
of the galaxy tidal threshing \citep{BCD01}.

At the same time, on the metallicity-velocity dispersion plot
(Fig.~\ref{figSLZ}, top panel) the loci of UCDs practically coincide with
those of dEs. We consider this as an argument for the scenario of tidal
threshing of dE,Ns as a way to create UCDs. In this
case, a velocity dispersion of a compact nucleus will not change very
strongly, while the total luminosity of a progenitor will drop by several
magnitudes. 

For comparison with massive GCs \citep{MvdM05} we chose a subsample of
Galactic GCs with available measurements of velocity dispersions. Many GCs
follow the behaviour of early-type galaxies. The three strongest outliers,
namely NGC~104, NGC~6388, and NGC~6441, similarly to UCDs, reside
significantly above the sequence of early-type galaxies on the metallicity
-- luminosity plot (Fig~\ref{figSLZ}). It is remarkable that the latter two
exhibit direct evidences of multiple stellar populations \citep{Piotto08}. 
Among the three other GCs (NGC~1851, NGC~2808, $\omega$~Cen) demonstrating
composite stellar populations only $\omega$~Cen follows exactly the trend
defined by early-type galaxies. We notice, that the third strongest outlier,
47~Tuc (NGC~104), being at least as massive as NGC~6388 does not have an
evident double main sequence \citep{Piotto08}.

In the frame of the tidal stripping scenario, we can also propose an
explanation for the large spread of UCD metallicities ($-$0.25 to $-$0.93)
on the [Fe/H]~vs.~$\sigma$ plot. It may be a superposition of the two
factors: (1) the relatively high spread of metallicities of dE progenitors
of UCDs due to their own environmentally-driven evolution (see discussion in
\citealp{Chil08A496}); (2) different conditions during the tidal
stripping, which may lead to some changes of the velocity dispersion
values compared to the progenitors.

\subsection{Comparison of dE,N nuclei and UCDs}

\begin{figure}
\includegraphics[width=0.95\hsize]{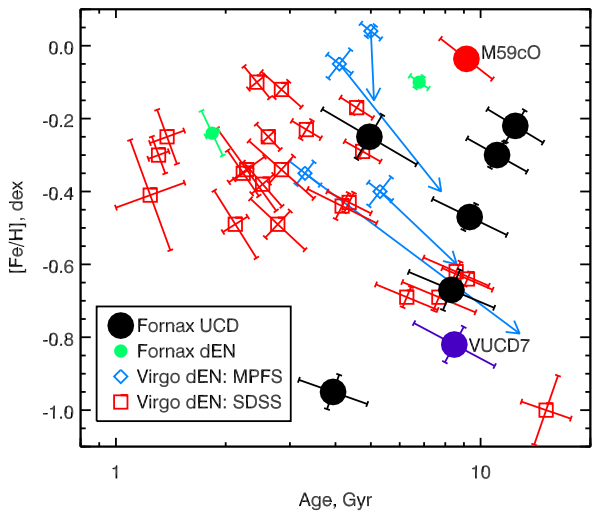}
\caption{Comparison of ages and metallicities of UCDs with a sample of
Virgo cluster dE,N galaxies from SDSS.\label{figtZ}}
\end{figure}

In Fig.~\ref{figtZ} we compare ages and metallicities of the 6 Fornax cluster
UCDs with nuclei of dE,Ns in the Fornax (2 objects, this study) and Virgo
(26 objects) clusters, transitional cE/UCD object M59cO \citep{CM08} and 
VUCD7, another UCD in the Virgo cluster. Stellar population
parameters for 22 Virgo cluster galaxies (shown in red), as well as for
VUCD7 and M59cO are obtained by analysing SDSS DR6 spectra. For the four
remaining Virgo dE,Ns shown in light blue we used the results based on the
3D spectroscopic observations presented in \cite{CPSA07,CSAP07}:
diamonds with error-bars correspond to the ages and metallicities of the
nuclei and the blue vectors point to the parameters of the ``main bodies''
of the galaxies.

Apart from the 2 intermediate age objects (UCD4 and UCD5), all UCDs are old,
at the same time spanning a large range of metallicities. Most of the dE,Ns
exhibit considerably younger stellar populations than UCDs. However,
there is a number of old dE,N nuclei (including FCC~182) with ages
comparable to those of UCDs. A scenario, assuming that dE,N nuclei are
results of repeated or extended star formation episodes in the dE centres,
leading to metal enrichment, can explain the
observed quite high metallicities of
dE,N nuclei.

\subsection{Origin of UCD galaxies}

The stellar population parameters obtained by our SSP fitting, 
namely metallicities higher than
$-$1.0 dex, allow us to exclude the scenario of evolving primordial density
fluctuations \citep{PDGJ01}, because one would expect much lower
metallicities for objects at this mass range.

Low dark matter content leaves space for all remaining channels of the UCD
formation: \cite{BCDS03} showed that in case of dE stripping (``threshing'')
the progenitor's nucleus must not be dark matter dominated; the two other
alternatives, GC merging and formation of UCDs as tidal superclusters,
assume zero dark matter content.

However, the scenario of merging GCs \citep{MHI02} fails to explain why we
do not observe metal-poor UCDs. It is known that GCs exhibit a dichotomy in
the metallicity distribution (e.g. review by \citealp{BS06}), but observed
UCDs correspond only to metal-rich GCs. Why don't metal-poor GCs merge?
Moreover, in case of a merger of metal-poor and metal-rich GCs of the same
mass, the resulting luminosity-weighted metallicity in our wavelength range
will be lower than the mean value, because metal-poor stellar populations
have lower $M/L$ ratios than metal-rich ones. Composite stellar
populations observed in massive Galactic GCs \citep{Piotto08} comprise two
to four SSPs, sometimes significantly different in metallicities, which is
evident from deep colour-magnitude diagrams. These objects look good
candidates for the GC merger scenario. In addition, we do see metal-poor
``composite'' GCs (NGC~1851, NGC~2808, $\omega$~Cen). Indeed, they are an
order of magnitude (except $\omega$~Cen, where at least four SSPs are
evident) fainter than the UCDs we are discussing here.

On the statistical basis, UCDs are too frequent to be representatives
of the high-luminosity end of the GC luminosity function (GCLF). After the
initial discussion in \cite{MHI02} a significant number of UCDs was
discovered. The extrapolation of the GCLF (see e.g. \citealp{ML07}) towards
bright objects results in the statistical over-population of $M_V < -11$
objects. Although, the exact value depends on the adopted GCLF parameters
and representation (i.e. Gaussian or $t_5$), we consider this fact
important, making this channel of UCD formation scarcely probable.

The old ages of most UCDs, compared to dE,N nuclei, suggest that if we
consider the dE,N tidal stripping scenario, it must have happened a long time
ago. However, there is a difficulty in explaining the formation of the most
metal-rich UCDs, because 8--10~Gyr ago dE,N nuclei must have been less metal
rich than presently observed. A possible explanation is a tidal stripping of
the most massive dE representatives (dE/E transitional objects) such as
IC~3653 or FCC~182. 

Another possibility is to create them as stellar superclusters (Fellhauer \&
Kroupa 2005) during interactions of massive galaxies, in this case
metal-rich population formed from the metal-rich gas of the progenitors 
will be observed in the UCDs. Both
scenarii are compatible with low dark matter content. A possible 
diagnosis is to measure $\alpha$/Fe abundance ratios: populations formed in a short
and intense star formation episode will be $\alpha$-overabundant (e.g.
\citealp{Matteucci94}).
With the present low S/N UCD data we are not able to carry out this test.

Finally, we are left with the two alternatives of UCD formation: UCDs with
low metallicities ($[\mbox{Fe/H}] < -0.5$~dex) are in favour of dE,N
tidal stripping, while tidally created superclusters better explain metal-rich
UCDs. At present we cannot exclude the diversity of the UCD origin suggested by 
\cite{MHIJ06}.

\section*{Acknowledgments}
We thank participants of the ``Nuclear Star Clusters Across the Hubble
Sequence'' workshop for fruitful discussions of the preliminary results, S.
Mieske for useful advices and discussions of UCD origin, our anonymous
referee for valuable comments, P. Di Matteo for providing a link to the
presentation of G.~Piotto. Special thanks to Gary Mamon for the critical
reading of the manuscript. GB is supported at he IAA/CSIC by an I3P
contract (I3P-PC2005-F) funded by the European Social Fund, with additional
support by DGI grant AYA 2005-07516-C02-01 and the Junta de Andaluc\'\i a.

\bibliographystyle{mn2e}
\bibliography{UCD_Fornax}

\label{lastpage}

\end{document}